\documentclass[conference,a4paper]{IEEEtran}
\usepackage{xcolor}
\usepackage{balance}

\ifCLASSINFOpdf
  \usepackage[pdftex]{graphicx}
\else
  \usepackage[dvips]{graphicx}
\fi
\usepackage{epstopdf}
\usepackage{graphicx} 
\usepackage{layout}
\usepackage{array}
\usepackage{amsmath}
\usepackage{amssymb}
\usepackage{tikz}
\usepackage{pgfplots}

\usepackage{amsmath}
\usepackage{mathtools}
\DeclarePairedDelimiter\norm{\lVert}{\rVert}
\ifCLASSOPTIONcompsoc
 \usepackage[caption=false,font=normalsize,labelfont=sf,textfont=sf]{subfig}
\else
 \usepackage[caption=false,font=footnotesize]{subfig}
\fi
\begin{document}
%
\title{A beamforming approach to the self-calibration of phased arrays}

\author{\IEEEauthorblockN{
Quentin Gueuning\IEEEauthorrefmark{1}, Anthony K. Brown\IEEEauthorrefmark{1}\IEEEauthorrefmark{2}, Christophe Craeye\IEEEauthorrefmark{3} and  
Eloy de Lera Acedo\IEEEauthorrefmark{1}  
}                                
\\
\IEEEauthorblockA{\IEEEauthorrefmark{1}
Cavendish Laboratory, University of Cambridge, Cambridge, UK}
\IEEEauthorblockA{\IEEEauthorrefmark{2}
Department of Electrical and Electronic Engineering, University of Manchester, Manchester, UK}
\IEEEauthorblockA{\IEEEauthorrefmark{3}
ICTEAM institute, Universit\' e Catholique de Louvain, Louvain-La-Neuve, Belgium} 

\IEEEauthorblockA{qdg20@cam.ac.uk}
}

\maketitle

\begin{abstract}
In this paper, we propose a beamforming method for the calibration of the direction-independent gain of the analog chains of aperture arrays. The gain estimates are obtained by cross-correlating the output voltage of each antenna with a voltage beamformed using the other antennas of the array. When the beamforming weights are equal to the average cross-correlated power, a relation is drawn with the StEFCal algorithm. An example illustrates this approach for few point sources and a $256$-element array.
\end{abstract}

\vskip0.5\baselineskip

%
%

\section{Introduction}
The calibration of the aperture array stations of the Square Kilometer Array (SKA) telescope \cite{SKA} and of its precursor and pathfinder, the Murchison Widefield Array (MWA) \cite{MWA} and the Low-Frequency Array (LOFAR) \cite{LOFAR}, is essential to achieve their expected nominal performance. The increased bandwidth and sensitivity, which has led to large irregular arrays of wideband antennas, inherently renders the station calibration more challenging due to complex coupling effects \cite{Ha}, each element being in a different local environment, and a higher data throughput \cite{Mitchell}. 

The SKA1-low instrument will consist of $512$ stations of $256$ elements \cite{SKA} covering the $50$-$350$ MHz band. The analog chain associated to each log-periodic antennas \cite{Eloy} will include low-noise amplifiers and other analog components connected via coaxial cables or optical fibers to remote analog-to-digital converters. The station beamforming will be performed digitally and thus relies on an accurate calibration of the analog chain. Given the limited storage capabilities of the digital backend and the fast time variability of the gains accounting for the propagation of the signals through the analog chain, one currently considers real-time calibrations \cite{Mitchell}.

Self-calibration \cite{Cornwell} is a calibration technique which uses a model of the sky and of the embedded element patterns (EEP) to solve for the gains.
Most algorithms implementing self-calibration formulate the problem as a quadratic system of equations based on a correlation matrix containing all the pairs of cross-correlation between the antenna voltages. For a station with many elements, those approaches are computationnally expensive since they scale at least quadratically w.r.t. the number of antennas \cite{Salvini}, i.e. $O(N^2)$ for $N$ antennas. Recently, \cite{Wijnholds} proposed a promising technique, called self-holography, which exhibits a linear complexity $O(N)$ and basically consists of cross-correlating each antenna voltage with the signal from a single beam formed by the same station pointing towards strong sources. Overall, beamforming approaches may help to decrease the computational complexity of the calibration since they avoid the construction of the correlation matrix. 

This paper presents a method for the gain calibration problem based on cross-correlating each antenna voltage with a voltage beamformed from the other antennas. The difference with \cite{Wijnholds} is that the beamforming weights, and thus the beams, are allowed to vary for each cross-correlation. This formulation should be well-suited to further accelerations using fast beamforming methods. We will then analyse the convergence and the accuracy of the numerical method for a simple scenario with few point sources and an irregular array similar to the SKA Aperture Array Verification System 1 (AAVS1) \cite{Ha}.

\section{Beamforming-based gain estimation}
The voltage $v_i(t)$ of antenna $i$ measured at the input port of the receiver can be decomposed into
\begin{align} \label{eq:signal}
v_i(t) = g_i s_i(t) + n_i(t)
\end{align}
where $g_i$ is a complex-valued gain accounting for the ith receiving path, $s_i(t)$ is the signal coming from the sky and $n_i(t)$ is the noise produced by the receiving system.
The signals are filtered to a bandwidth $B$ and sampled in time at the Nyquist frequency $1/(2B)$ for a duration $T$.
The response of the overall system is assumed frequency-flat over the bandwidth $B$. 

\begin{figure}[t]
\centering
\begin{tikzpicture}
\draw [black,very thick] (0,0.0) -- (0,0.5);
\draw [black,very thick] (0,0.5) -- (0.5,1.0);
\draw [black,very thick] (0,0.5) -- (-0.5,1.0);

\draw [black,very thick] (-1.3,0.0) -- (-1.3,0.5);
\draw [black,very thick] (-1.3,0.5) -- (0.5-1.3,1.0);
\draw [black,very thick] (-1.3,0.5) -- (-0.5-1.3,1.0);

\draw [black,very thick] (-3.0,0.0) -- (-3.0,0.5);
\draw [black,very thick] (-3.0,0.5) -- (0.5-3.0,1.0);
\draw [black,very thick] (-3.0,0.5) -- (-0.5-3.0,1.0);

\draw [black,very thick] (-4.8,0.0) -- (-4.8,0.5);
\draw [black,very thick] (-4.8,0.5) -- (0.5-4.8,1.0);
\draw [black,very thick] (-4.8,0.5) -- (-0.5-4.8,1.0);

\draw [black,very thick] (-6.2,0.0) -- (-6.2,0.5);
\draw [black,very thick] (-6.2,0.5) -- (0.5-6.2,1.0);
\draw [black,very thick] (-6.2,0.5) -- (-0.5-6.2,1.0);

\draw [black] (-6.2,-0.0) -- (-6.2,-0.5) ;
\draw [black] (-3.0,0.0) -- (-3.0,-0.5) ;
\draw [black] (0,0.0) -- (0,-0.5) ;
\draw [black] (-1.3,0.0) -- (-1.3,-0.5) ;

\draw [black] (-6.2,-0.5) -- (-3.5-1.2,-1.25) ;
\draw [black] (-1.3,-0.5) -- (-1.5-1.2,-1.25) ;
\draw [black]  (-3.0,-0.5)-- (-2.5-1.2,-1.25);
\draw [black] (0,-0.5) -- (-0.5-1.2,-1.25);

\draw [black] (-3.5-1.2,-1.25)--(-3.5-1.2,-2.0) ;
\draw [black] (-1.5-1.2,-1.25)--(-1.5-1.2,-2.0);
\draw [black] (-2.5-1.2,-1.25)--(-2.5-1.2,-2.0);
\draw [black] (-0.5-1.2,-1.25)--(-0.5-1.2,-2.0);

\draw [black] (-3.5-1.2,-2.0)-- (-2.0-1.2,-3.0) ;
\draw [black] (-1.5-1.2,-2.0)-- (-2.0-1.2,-3.0) ;
\draw [black] (-2.5-1.2,-2.0)-- (-2.0-1.2,-3.0) ;
\draw [black] (-0.5-1.2,-2.0)-- (-2.0-1.2,-3.0) ;

\draw [black] (-4.8,0.0) -- (-4.8,-0.5);
\draw [black] (-4.8,-0.5) -- (-6.2,-1.25);
\draw [black,->] (-6.2,-1.25) -- (-6.2,-4.25);
\node[text width=0.5cm,align=left] at (-6.2,-4.5)
{$v_2(t)$};

\draw [ black,fill=white] (-3.5-1.2,-2.0)  circle (0.15cm);
\draw [black] (-3.5-0.1061-1.2,-2.0+0.1061) -- (-3.5+0.1061-1.2,-2.0-0.1061);
\draw [black] (-3.5+0.1061-1.2,-2.0+0.1061) -- (-3.5-0.1061-1.2,-2.0-0.1061);
\draw [ black,fill=white]  (-1.5-1.2,-2.0) circle (0.15cm);
\draw [black] (-1.5-0.1061-1.2,-2.0+0.1061) -- (-1.5+0.1061-1.2,-2.0-0.1061);
\draw [black] (-1.5+0.1061-1.2,-2.0+0.1061) -- (-1.5-0.1061-1.2,-2.0-0.1061);
\draw [ black,fill=white] (-2.5-1.2,-2.0) circle (0.15cm);
\draw [black] (-2.5-0.1061-1.2,-2.0+0.1061) -- (-2.5+0.1061-1.2,-2.0-0.1061);
\draw [black] (-2.5+0.1061-1.2,-2.0+0.1061) -- (-2.5-0.1061-1.2,-2.0-0.1061);
\draw [ black,fill=white] (-0.5-1.2,-2.0) circle (0.15cm);
\draw [black] (-0.5-0.1061-1.2,-2.0+0.1061) -- (-0.5+0.1061-1.2,-2.0-0.1061);
\draw [black] (-0.5+0.1061-1.2,-2.0+0.1061) -- (-0.5-0.1061-1.2,-2.0-0.1061);

\node[text width=0.5cm,align=left] at (-5.0,-3.5)
{beamformer};

\draw [draw=black] (-1.0,-3.75) rectangle (-5.45,-1.25);

\node[text width=0.5cm,align=left] at (-3.5-1.6,-1.6)
{$\hat{w}_{21}$};
\node[text width=0.5cm,align=left] at (-2.5-1.6,-1.6)
{$\hat{w}_{23}$};
\node[text width=0.5cm,align=left] at (-1.5-1.6,-1.6) 
{$\hat{w}_{24}$};
\node[text width=0.5cm,align=left] at (-0.5-1.6,-1.6)
{$\hat{w}_{25}$};

\draw [black,dotted] (-6.6,-0.5) -- (0.4,-0.5);

\node[text width=0.5cm,align=left] at (-6.8,-0.25)
{$v_1(t)$};
\node[text width=0.5cm,align=left] at (-5.4,-0.25)
{$v_2(t)$};
\node[text width=0.5cm,align=left] at (-1.9,-0.25)
{$v_4(t)$};
\node[text width=0.5cm,align=left] at (-3.6,-0.25) 
{$v_3(t)$};
\node[text width=0.5cm,align=left] at (-0.6,-0.25)
{$v_5(t)$};

\draw [ black,fill=white] (-2.0-1.2,-3.0) circle (0.15cm);
\draw [black] (-2.0-0.15-1.2,-3.0) -- (-2.0+0.15-1.2,-3.0);
\draw [black] (-2.0-1.2,-3.0-0.15) -- (-2.0-1.2,-3.0+0.15);

\draw [black,->] (-2.0-1.2,-3.0-0.15) -- (-2.0-1.2,-4.25);
\node[text width=0.5cm,align=left] at (-3.25,-4.5)
{$b_2(t)$};

\draw[scale=1,domain=1:179,variable=\x,black,dashed] plot ({1.5*sin(\x)*sin(\x)*cos(\x)-4.8},{1.5*sin(\x)*sin(\x)*sin(\x)+1}) ;
\draw[scale=1,domain=1:179,samples=104,variable=\x,black,dashed] plot ({0.08*sin(8*180/3.14*(cos(\x)+0.2))/(cos(\x)+0.2)*sin(8*180/3.14*(cos(\x)+0.2))/(cos(\x)+0.2)*cos(\x)-3.1},{0.08*sin(8*180/3.14*(cos(\x)+0.2))/(cos(\x)+0.2)*sin(8*180/3.14*(cos(\x)+0.2))/(cos(\x)+0.2)*sin(\x)+1}) ;

\end{tikzpicture}
\caption{Illustration of the cross-correlation between the voltage $v_2$ and the beamformed voltage $b_2$ used to determine the gain $g_2$. The dotted line and the dashed lines represent, respectively, the ADC output and the radiation patterns associated to $v_2$ and $b_2$.}
\label{fig:beamformed}
\end{figure}
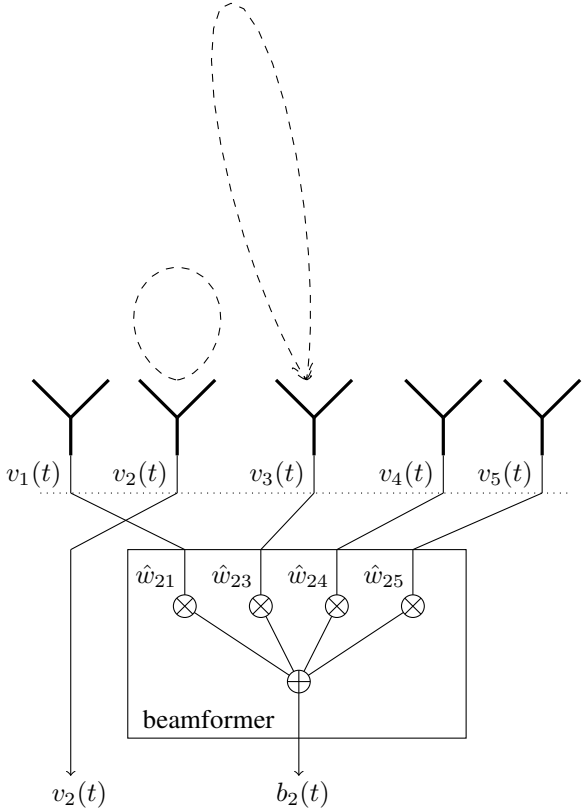

The calibration technique, developed next, iteratively finds the unknown gains $g_i$ by cross-correlating the voltage of a single element $i$ with a voltage beamformed using all the other elements, as illustrated in Fig.~\ref{fig:beamformed}. At each iteration, the gains of the beamformed array are assumed perfectly known s.t. only $g_i$ needs to be found. Intuitively, this approximation is relevant when the errors on the gains average out during the beamforming operation.

The cross-correlation product between the voltage $ v_i(t) $ of antenna $i$ and the voltage $ b_i(t) $ at the output of the array beamformer is measured by
\begin{align} \label{eq:timeavcorr}
R_{bi} = \frac{1}{M} \ \sum_{m} v_i(t_m)   b_i(t_m)^\star
\end{align}
where $\star$ stands for complex conjugate and $M = 2BT$ is the number of time samples $t_m = m/(2B)$. As illustrated in Fig.~\ref{fig:beamformed} for $i=2$, the beamformed voltage is computed with
\begin{align} \label{eq:beamforming}
b_i(t) = \hat{\mathbf{w}}_{i} \cdot \mathbf{v}(t)
\end{align}
where $\hat{\mathbf{w}}_{i}$ is a normalized vector containing the beamforming weights $\hat{w}_{ij}$ and the antenna voltages have been stacked into the vector $\mathbf{v}(t)$.

Assuming ergodicity and uncorrelated noises $n_i$, the mean over time of the
correlation product $P_{ij}$  is equal to the ensemble average $\langle v_i v_j^\star \rangle$ and is given by
\begin{equation} \label{eq:corrtemp}
P_{ij} = k_b  \left(g_i T_{ij} g_j ^\star +  T_{ni}  \delta_{ ij } \right) B
\end{equation}
where $k_b$ is the Boltzmann's constant, $\delta _{ij}$ is the Kronecker's delta, $T_{ij}  =  \langle s_i s_j^\star  \rangle /(k_b B)$ and $T_{ni} = \langle n_i n_i^\star \rangle/(k_b B) $ are, respectively, the correlated antenna temperature and the noise temperature. In self-calibration, the correlation product $\langle s_i s_j^\star  \rangle$ is known and can be evaluated numerically from models of the EEPs and of the source intensities \cite{Clark}.

Using \eqref{eq:beamforming} and \eqref{eq:corrtemp}, the correlated power $P_{bi} = \langle v_i b_i^\star \rangle $ is expressed as 
\begin{equation} \label{eq:meancross}
 P_{bi}=  k_b  \left(g_i  T_{bi} + \hat{w}^\star _{ii}  T_{ni} \right) B
\end{equation}
where $T_{bi} = \mathbf{\hat{w}}_{i}^\star \cdot (\mathbf{g}^\star  \odot \mathbf{T}_{i})$ is the beamformed array temperature with the element-wise product $\odot$, the gain vector $\mathbf{g}$  and  the vector $\mathbf{T}_i$ standing for the ith row of the matrix $\mathbf{T}$ which contains all the antenna temperatures $T_{ij}$.
Assuming that a guess for $T_{bi}$ has been obtained at the previous iteration and that the signals $s_i$ are ergodic, we can make the approximation $P_{bi} \approx R_{bi}$ and isolate the gain $g_i$ in \eqref{eq:meancross} to have
\begin{equation}\label{eq:estimator}
\tilde{g}_i  = \frac{R_{bi}- \hat{w}^\star_{ii} k_b T_{ni} B  } {k_b T_{bi} B}
\end{equation}
The gain estimation with \eqref{eq:estimator} can be viewed as two-step method with noise extraction followed by rescaling. It requires the knowledge of the noise temperature $T_{ni}$. When no noise information is available, one can cancel the auto-correlations in \eqref{eq:timeavcorr} by imposing  $\hat{w}_{ii} = 0$.

Next, if we insert \eqref{eq:timeavcorr} into \eqref{eq:estimator} and define the vector $\tilde{\mathbf{g}}$ containing the gain estimates $\tilde{g}_i$, the following matrix form is obtained
\begin{equation}\label{eq:estimatormat}
\tilde{\mathbf{g}}  = \left( \sum_m \mathbf{v}(t_m) \odot \left(\mathbf{W}  \mathbf{v}(t_m)\right)^\star \right) \oslash (Mk_b \mathbf{T}_b B)
\end{equation}
where  $\oslash$ denote the element-wise division, $\mathbf{W}$ is a matrix whose lines are the weights $\mathbf{\hat{w}}_i$  and $\mathbf{T}_b$ is a vector containing the array temperatures $T_{bi}$. It is clear that a brute force evaluation of the matrix-vector product in \eqref{eq:estimatormat} will lead to $O(N^2)$ complexity per time sample. However,  it is expected that fast beamforming methods for irregular array will help to decrease the computational burden.

In the following example, we will only consider the beamforming weights 
\begin{equation}\label{eq:Stefcalweights}
\mathbf{\hat{w}}_i = \frac{\mathbf{g}^\star  \odot \mathbf{T}_{i}}{\norm{\mathbf{g}^\star  \odot \mathbf{T}_{i}}}
\end{equation}
which maximize the average cross-correlated power $P_{bi}$ in \eqref{eq:meancross}. It can be proven that the particular choice of weights \eqref{eq:Stefcalweights} leads to gain estimates $\tilde{\mathbf{g}}$  equal to those obtained after a single iteration of the StEFCal algorithm. Indeed, by comparing \eqref{eq:estimatormat} with the relation (12) in \cite{Salvini} using the same notations, one can easily identify the equivalences $\mathbf{\hat{w}}_p = \mathbf{Z}_{:,p} / \norm{\mathbf{Z}_{:,p}}$ and $ k_b T_{bp} B = \mathbf{Z}_{:,p}^H \cdot \mathbf{Z}_{:,p}$.

\section{Numerical example}

\begin{figure} [t]
\centering
 \includegraphics[trim=0.0cm 0.0cm 0.0cm 0.0cm,clip,width=9.0cm,height=6.8cm]{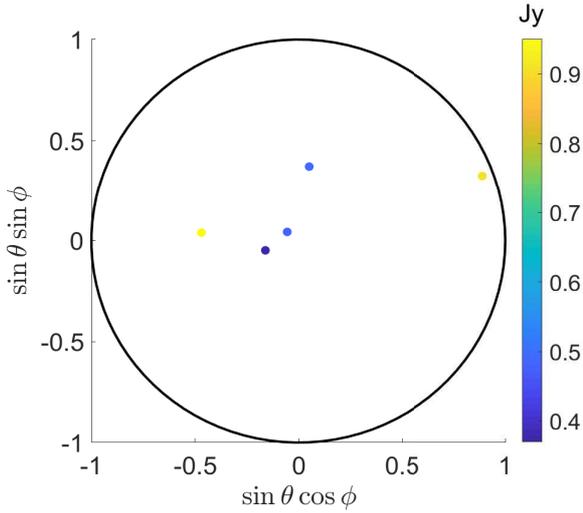}
 \caption{Intensity of the point-sources w.r.t. the spherical coordinates $(\theta,\phi)$.}
 \label{fig:sky}
\end{figure}

We will now apply the proposed method to calibrate an irregular array of $256$ SKALA4 antennas \cite{Eloy} with a random layout similar to the AAVS1 array \cite{Ha}. The noise temperature of the system is fixed to $T_{ni} = 200$ K for each antenna. As shown in Fig.~\ref{fig:sky}, we have created a sky map composed of $5$ unresolved and unpolarized sources of intensities randomly distributed between $0.1$ and $1 $ Jy. Both the signals $s_i (t)$ and $n_i (t)$ are Gaussian random variables filtered to a bandwidth $B=1$ MHz around the center frequency $f = 148 $ MHz. The cross-correlations $R_{bi}$ in \eqref{eq:estimator} are evaluated with an integration time $T = 0.1 s$ leading to $M = 2 \times 10^5$ time samples. 
The EEP of each antennas, necessary to compute the correlated temperatures $T_{ij}$ in \eqref{eq:corrtemp},  have been simulated with the HARP software \cite{Ha} and takes into account the mutual coupling effect. The amplitude and phase of the true gains $g_i$ are randomly distributed within the intervals $[0,2]$ and $[0,\pi]$, respectively. The gain estimates at the initial step are set to $\tilde{g}_i = 1$. The method is stopped after $3$ iterations. Each iteration takes a full timeslot. Hence, the duration of the calibration is $0.3 s$.

\begin{figure} [t]
\centering
 \includegraphics[trim=0.0cm 0.0cm 0.0cm 0.0cm,clip,width=9.0cm,height=6.8cm]{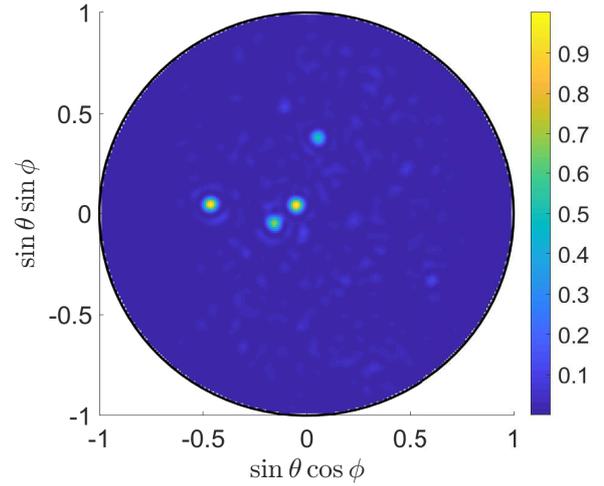}
 \caption{Beam formed at the third iteration when cross-correlating with a particular element.}
 \label{fig:beam}
\end{figure}

\begin{figure} [t]
\centering
 \includegraphics[trim=0.0cm 0.0cm 0.0cm 0.0cm,clip,width=9.0cm,height=7.0cm]{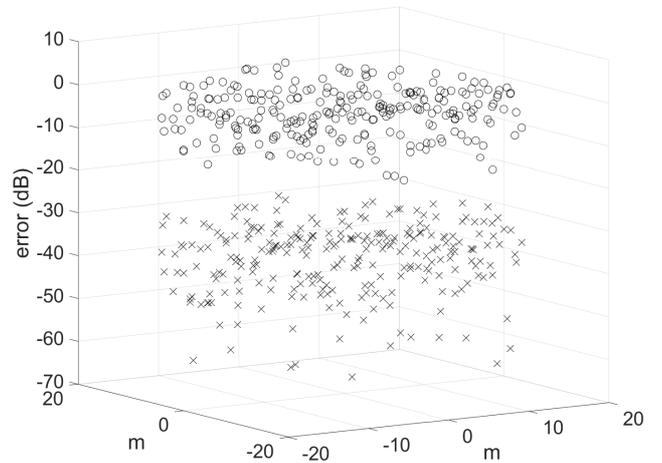}
 \caption{Relative error on the gain estimates initially (circles) and after the third iteration (cross) with the beamforming method.}
 \label{fig:errors}
\end{figure}

One can observe in Fig.~\ref{fig:beam} that the beam formed with the weights \eqref{eq:Stefcalweights} at the third iteration is producing hot spots in the direction of the point sources. The source close to the horizon is an exception and has been weighted down by the EEPs since the field of view of the SKALA antenna is limited. In Fig.~\ref{fig:errors}, one can see that the relative error, computed with $\epsilon= |\tilde{g}_i-g_i|/|g_i|$,  has decreased from $0$ dB for the initial guess to around $-30$ dB after the third iteration. This thus provides a simple numerical validation of the proposed self-calibration technique.

\section{Conclusion}

We have presented and validated a beamforming method based on the computation of the cross-correlation between each antenna voltage and the voltage formed by a beam using the other antennas of the station. When the beamforming weights match the average antenna cross-correlation, the method is equivalent to a single iteration of StEFCal algorithm. This thus ensures the convergence and the efficiency of the gain estimates for most practical cases.

In the way that fast methods have accelerated matrix-vector products in computational electromagnetics, fast beamforming methods, for instance using the Non-equispaced Fast Fourier Transform \cite{Dutt}, will probably help to bring down the computationnal complexity from  $N^2$ for classical approaches to $N\log _2 N$ per time sample.


\end{document}